\documentstyle[12pt,axodraw]{article}
\begin{document}
\baselineskip=6.0mm
\begin{titlepage}
%%%%%%%%%%%%%%%%%%%%%%%%%%%%%%%%%%%%%%%%%%%%%%%%%
\begin{flushright}
  KOBE-TH-98-06\\
\end{flushright}
%%%%%%%%%%%%%%%%%%%%%%%%%%%%%%%%%%%%%%%%%%%%%%%%%%%
\vspace{2.3cm}
\centerline{{\large{\bf Lepton-Flavour Violation}}}
\centerline{{\large{\bf in}}}
\centerline{{\large{\bf Ordinary and Supersymmetric Grand Unified
Theories}}}
\par
\par
\par\bigskip
\par\bigskip
\par\bigskip
\par\bigskip
\par\bigskip
\renewcommand{\thefootnote}{\fnsymbol{footnote}}
\centerline{{\bf C.S.
Lim}$^{(a)}$\footnote[1]{e-mail:lim@octopus.phys.sci.kobe-u.ac.jp}
 and {\bf Bungo
Taga}$^{(b)}$\footnote[2]{e-mail:taga@octopus.phys.sci.kobe-u.ac.jp}}
\par
\par\bigskip
\par\bigskip
\centerline{$^{(a)}$ Department of Physics, Kobe University, Nada, Kobe 
657-8501,
Japan}
\centerline{$^{(b)}$ Graduate School of Science and Technology, Kobe
University, Kobe 657-8501,
Japan}
\par
\par\bigskip
\par\bigskip
\par\bigskip
\par\bigskip
\par\bigskip
%\centerline{\today }\par
\par\bigskip
\par\bigskip
\par\bigskip
\par\bigskip
\centerline{{\bf Abstract}}\par

By an explicit calculation we show that in ordinary SU(5) logarithmic divergence in the amplitude of $\mu \rightarrow e\gamma$ cancels among  diagrams and remaining finite part is suppressed by  at least $1/M_{GUT}^2$. In SUSY SU(5), when the effect of flavour changing wave function renormalization is taken into account  such logarithmic correction disappears, provided a condition is met among SUSY  breaking masses. In SUGRA-inspired SUSY GUT the remaining logarithmic effect is argued not to be taken as a prediction of the theory.

\par\bigskip
\par\bigskip
\par\bigskip
\par\bigskip
\par\bigskip
\par\bigskip
\par\bigskip

%\noindent September 1995
\end{titlepage}
\newpage

Flavour changing neutral current (FCNC) processes have long served as a good thinking ground and a good experimental probe of ``new physics" in each stage of the development of high energy physics. The violation of flavour symmetry in the lepton sector, ``lepton flavour violation (LFV)", is of special interest to us. FCNC processes due to LFV  are strictly forbidden in the Standard Model with massless neutrinos. Thus LFV is a very clean signal of new physics, if it  exists. Our main interest in this paper is in $\mu \rightarrow e\gamma$, whose present experimental bound is $Br(\mu \rightarrow e\gamma) < 4.9\times 10^{-11}$.

In this context, a very interesting claim has been made by Barbieri - Hall and collaborators that in SUSY GUT models ``sizable" rates of $\mu \rightarrow e\gamma$ are expected, contrary to ordinary expectation \cite{Barbieri}. Such claim has been followed by many works calculating the rate in a few GUT theories \cite{Muegamma}. They work in R-conserving SUGRA-inspired SUSY GUT models. The crucial observation there is that the large flavour violation due to top quark Yukawa coupling combined  with GUT interaction yields sizable non-universal or LFV renormalization group effects on the SUSY breaking slepton masses. They cause the discrepancy between mass matrices of charged leptons and their superpartners. Thus super-GIM mechanism \cite{Gatto} is no longer valid and  photino-exchange, for instance, leads to $\mu \rightarrow e\gamma$ , at the rates which is not  so far from the experimental upper bound. While their result \cite{Barbieri}is quite impressive, it seems to be different from what we usually expect in 
the following sense: (i) The reason they got sizable rates is the appearance of the LFV slepton masses due to renormalization group effect, which are roughly proportional to $ln \frac{\Lambda}{M_{GUT}}$, instead of $\frac{m_{\mu}^{2}}{M_{GUT}^2}$. The difference is tremendous ! The U.V.cutoff $\Lambda$ was taken to be the Planck scale, $\Lambda = M_{pl}$.  This in turn means that  logarithmic divergence remains in the quantum corrections to LFV processes in SUSY GUT, and seems to contradict with what we have seen in the standard model \cite{Gaillard}, \cite{Inami} and what we expect in ordinary non-SUSY SU(5) discussed below, where the amplitudes of FCNC processes are automatically finite simply because the relevant operators do not  exist in the original lagrangian: natural flavour conservation holds. (ii)  Next, the $ln \frac{\Lambda}{M_{GUT}}$ contribution clearly shows that GUT particles  do not decouple from the low energy process, in contradiction with a general argument for the decoupling of particle 
with GUT scale masses. In the  SU(2)$\times$U(I) theory,  as the low energy effective theory of ordinary GUT,  LFV should be described by induced SU(2)$\times$U(I) invariant irrelevant operators simply because LFV does not exist in the original effective theory. The coefficient should be suppressed by the inverse powers of $M_{GUT}$, as LFV never appears without GUT interactions. This proves the decoupling.

The purpose of our work \cite{Lim} is to clarify whether the interesting features claimed in Ref.\cite{Barbieri} are natural consequences of SUSY GUT theories. We will compare LFV in ordinary  and SUSY SU(5) GUT, in order to see how SUSY can be essential in getting the sizable effects.

One remark is in order. In our analysis a special attention will be paid on the effect of lepton flavour changing (or LFV) wave function renormalization, as  a similar effect played a central role in the cancellation of U.V. divergences in the standard model. More precisely, in the calculations of quark FCNC amplitudes in the standard model, both of flavour changing self-energy diagram (e.g. $\overline{s}i\partial_{\mu}\gamma^{\mu}b$) and flavour changing vertex diagram (e.g. $\overline{s}\gamma^{\mu}b A_{\mu}$, with $A_{\mu}$ being a photon field) have U.V. divergences, due to the hard flavour violation by Yukawa couplings in unphysical scalar exchange diagrams. What we find \cite{Inami},\cite{Botella}, however, is that there is an exact cancellation  of U.V. divergence between the diagrams where a photon is attached to the external legs of the  flavour changing self-energy diagram (``external leg correction" diagrams) and the flavour changing proper vertex diagram. Thus the final result is finite. This 
situation is easily understood in terms of operator language as follows. As stated above, after quantum corrections both the kinetic term of quarks and the minimal coupling term of photon are modified such that they have off-diagonal flavour changing pieces. Due to the $U(1)_{em}$ invariance, however, these corrections are not independent (``generalized Ward identity") and are described by a marginal operator with respect to a column vector of the quark bare fields $\psi_0$ as,
\begin{equation}
\overline{\psi_0}iD_{\mu}\gamma^{\mu} H \psi_0,
\end{equation}
where   $ H = H_R \frac{1 + \gamma_{5}}{2} +  H_L \frac{1-\gamma_{5}}{2}$ and the 3 $\times$ 3 hermitian matrices $H_R, H_L$ are generally non-diagonal. Then we can perform unitary transformations to the renormalized fields $\psi$ so that their kinetic term is flavour diagonal, $U_L^{\dag}H_LU_L = H_L^{diagonal}$, for instance. These unitary transformations, at the same time, diagonalize the photon vertex coming from the covariant derivative $D_{\mu}$. We may further perform the rescaling of the fields $\psi$ so that the kinetic term is proportional to  a unit matrix. Then another unitary transformation becomes possible to make the Yukawa coupling of the quarks flavour diagonal again, while keeping the form of the kinetic term. Thus FCNC disappears from the whole marginal operator, even after the quantum corrections, thus leading to the finiteness of FCNC amplitudes.

Now we are ready to discuss LFV in ordinary SU(5) GUT (The details are given in  Ref.\cite{Taga}.). In GUT theories,  even if neutrinos are massless, GUT interactions connecting charged leptons with quarks make LFV possible, though the rates are expected to be quite strongly suppressed by $1/M_{GUT}^4$: decoupling. We confirm this expectation by explicit calculations. As LFV becomes possible solely due to GUT interactions, we may focus only on the diagrams with exchanges of heavy particles with GUT mass scale $M_{GUT}$. We find that out of 14 Feynman diagrams, only the diagrams where a color triplet Higgs is exchanged give dominant contributions. Let $L_{eff}$ be an effective lagrangian, which is responsible for the $\mu \rightarrow e\gamma$ decay,
\begin{equation}
L_{eff} = c_{LFV}\times \overline{e}\sigma_{\mu\nu}(m_{\mu}\frac{1 -
\gamma_5}{2} +
m_{e}\frac{1 + \gamma_5}{2}) \mu \times F^{\mu\nu},
\end{equation}
where $ F^{\mu\nu}$ is the photon field strength. At the order of $1/M_{GUT}^2$  the coefficient function $c_{LFV}$ is given as
\begin{equation}
c_{LFV} = - \frac{\sqrt{6}}{16}\frac{g^3}{(4\pi)^2}
\frac{1}{M_h^2}(V_{KM}^{\dag})_{\mu
j}(V_{KM})_{je}\frac{m_{uj}^2}{M_W^2}(2ln\frac{m_{uj}^2}{M_h^2} +
\frac{15}{4}),
\end{equation}
where $M_h$ and $m_{uj}$ denote the masses of the color triplet Higgs $h$ and i-th up-type quark and the Kobayashi-Maskawa matrix $V_{KM}$ handles the LFV. Unfortunately, the resultant branching ratio is  almost nothing, being suppressed by $m_{\mu}^4/M_{h}^4 \sim m_{\mu}^4/M_{GUT}^4$. We have shown two things, i.e., (i) the log-divergence cancells out when the sum of all $h$-exchange diagrams is taken, and  (ii) the decoupling of GUT particles also holds; after the cancellation of the log-divergence and constant terms, the remaining amplitude is really suppressed at least by $m_{uj}^2/M_{GUT}^2$.

Let us now move on to the discussion of the LFV in SUSY SU(5) GUT. In this context, it has been claimed \cite{Barbieri} that the rate of $\mu \rightarrow e\gamma$ can be `sizable'. We, however, have already seen right above that such sizable effect does not appear in ordinary SU(5). Why $\mu \rightarrow e\gamma$ is enhanced once the theory is made supersymmetric ? Conceptually SUSY itself has nothing to do with flavour symmetry. More explicitly, as stated earlier, we have the following questions concerning the result; (i) Why does the logarithmic-divergence $ln \frac{\Lambda^2}{M_{GUT}^2} = ln \frac{M_{pl}^2}{M_{GUT}^2}$ remain ? (ii)Why is the suppression by a factor $(\frac{M_{W}^4}{M_{GUT}^4})$, i.e. the decoupling, absent ? We have shown that these are not the case in non-SUSY SU(5) GUT. We know that as an important  new feature of the SUSY GUT the soft SUSY breaking masses can be new source of flavour violation. We, however, still have the following questions; (iii) In Ref.\cite{Barbieri}, soft SUSY 
breaking masses have been assumed to be universal, being flavour independent, at the tree level and cannot be a new source of LFV by itself. Then why did such drastic change as $(\frac{M_{W}^2}{M_{GUT}^2})$ $\rightarrow ln \frac{M_{pl}^2}{M_{GUT}^2}$ become possible ? (iv) SUSY breaking terms are quite soft in the sense $M_{SUSY} \ll M_{GUT}, \ M_{pl}= \Lambda$. Then how can they drastically affect the U.V.-divergence, though one usually expects that soft breaking of some symmetry is harmless concerning the renormalizability of the theory constrained by the symmetry ?

  Since LFV in SUSY GUT is so interesting and important issue, to settle the above questions seems to be meaningful. We try to reanalyze the issue with the help of concrete computations. Our main interest is in the point whether or not the logarithmic divergence remains as a natural property of the theory and whether the general argument  supporting the decoupling of the effects due to  GUT interactions really breaks down. The model to work with is SUSY SU(5) GUT with explicit soft SUSY breaking terms. Following Ref.\cite{Barbieri}, we assume that bare SUSY breaking masses, the masses at the cutoff $\Lambda$, are flavour-independent. In particular, for right-handed charged sleptons,  they are given as $m_{0l}^2 (|\tilde{e}_R|^2 + |\tilde{\mu}_R|^2 + ...)$, where $m_{0l}$ denotes a universal bare SUSY breaking mass. In MSSM such flavour-independent masses will enable us to diagonalize both lepton and slepton mass matrices simultaneously (super GIM-mechanism \cite{Gatto}), and for massless neutrinos there will
 be no LFV, just as  in the standard model. As was correctly pointed out by Barbieri-Hall and  collaborators, the situation changes in SUSY SU(5), because of the presence of GUT interactions. The GUT interaction accompanied by large top Yukawa coupling $f_t$ which connects charged sleptons with stop $\tilde{t}$, for instance, can be new source of LFV. When combined with SUSY breaking, such effects lead to LFV SUSY breaking masses like $(f_t^2 ln \frac{\Lambda^2}{M_{GUT}^2} M_{SUSY}^2) \ \tilde{e}_R^{\ast}\tilde{\mu}_R$ at the 1-loop level. According to Ref.\cite{Barbieri}, the induced logarithmic correction to the slepton mass-squared term leads to $\mu \rightarrow e\gamma$ through the ordinary MSSM interactions, e.g. photino-exchange diagram, with a rate not suppressed by $1/M_{GUT}^4$. Thus actually  the $\mu \rightarrow e\gamma$ process is induced at 2-loop level in this scenario.

The most rigorous way to reach the rate of $\mu \rightarrow e\gamma$ is to directly calculate all possible 2-loop diagrams, which we would like to avoid. Instead, we may take the following approach to analyze the effect. First, we perform the path-integral from the cutoff $\Lambda$ to some scale $\mu$ , satisfying  $M_W \ll \mu \ll M_{GUT}$ (``Wilsonian renormalization"), at the 1-loop level. We thus obtain SU(3)$\times$SU(2)$\times$U(1) invariant effective low-energy ( $E \leq \mu$ ) lagrangian $L_{eff}$ with respect to light ($\ll M_{GUT}$) particles, which should be identified with MSSM.  Then, by using the induced LFV masses for charged sleptons in $L_{eff}$ the rate of $\mu \rightarrow e\gamma$  can be calculated just as in MSSM.  Therefore,  our focus should be on the point whether the (non-decoupling) lepton flavour changing logarithmic quantum correction ever appears in $L_{eff}$. The effective lagrangian can be decomposed into two parts, $L_{eff} = L_{rel} + L_{irrel}$, where $L_{rel}$ includes 
operators with mass dimension $d \le 4$, while $L_{irrel}$ denotes the set of irrelevant operators with $d > 4$. Since LFV stems solely from GUT particle exchanges, only their contributions to the LFV parts of $L_{eff}$ are considered in our analysis. Some remarks are in order; (a) Even if we get flavour changing slepton masses, they do not immediately lead to the presence of a net FCNC effect. As we have seen in the introductory argument, the effects of flavour changing wave-function renormalization should also be taken into accounts, and a machanism to cancel the logarithmic divergence is expected to work.  This possible cancellation mechanism does not seem to have been addressed in the previous analysis \cite{Barbieri}, \cite{Muegamma}. (b)  The GUT particle contributions to $L_{irrel}$ will be suppressed by the inverse powers of $M_{GUT}$. Thus only the contributions to $L_{rel}$ will be considered below. (c) Although they are ``soft", the SUSY breaking terms potentially affect operators with $d \leq 3$ in $L_{rel}$, but only up to $O(M_{SUSY}^2)$. For instance, the $M_{SUSY}^2$ insertion to the 1-loop diagram for LFV mass operator $\tilde{e}^{\ast}\tilde{\mu}$ yields log-divergence, while one more insertion of $M_{SUSY}^2$ will make the diagram finite, being suppressed by the inverse powers of $M_{GUT}$. (d) To get the flavour changing slepton masses, not only $M_{SUSY}^2$ but also  flavour violation are necessary. Only hard flavour violation due to the top quark Yukawa  coupling via the exchange of color triplet Higgs superfield will be important. The soft flavour violation due to the  mass-squared differences of up-type quarks contributes to the process only as a sort of irrelevant operator, since $M_{SUSY}^2$ insertion together with the insertion of Higgs doublet twice to provide the up-quark masses makes the operator higher-dimensional. 

Hence our task is to calculate the quantum corrections due to the exchange of colored Higgs superfield to the slepton part of $L_{rel}$ ( The quantum corrections to charged leptons are not independent of the supersymmetric terms of the slepton part, both being diagonalized symultaneously. Thus we need not calculate them independently). The relevant part of the calculated $L_{eff}$ (in momentum space) takes the following form;
\begin{eqnarray}
L_{eff}
&=&  \tilde{l}_i^{\ast}(p) [ (\delta_{ij} + a H_{ij} ) p^2 - (\delta_{ij} + 
b H_{ij}) m_{0l}^2 ] \tilde{l}_j(p)  \nonumber \\
&-&e \tilde{l}_i^{\ast}(p) (p + p')_{\mu} (\delta_{ij} + c H_{ij} )
\tilde{l}_j(p') \cdot A^{\mu}(q),
\end{eqnarray}
where $\tilde{l}_i = \tilde{e}_R, \tilde{\mu}_R$, etc., and $q = p - p'$. $A^{\mu}$ denotes a photon field. The $m_{0l}$ is the SUSY breaking mass for slepton, and $H_{ij} = f_{ti}f_{tj}^{\ast}$ with $ f_{t\mu} = g(\frac{m_t}{M_W})(V_{KM})_{ts}$, for instance. Let us note that the large top Yukawa coupling gives contributions only to the operators with respect to right-handed sleptons. Thus terms in the above equation are all chirality preserving, and the SUSY breaking mass-squared term with $m_{0l}^2$ should be treated on an equal footing with the self energy term accompanied by $p^2$. The coefficients $a,b$ and $c$ contain the results of 1-loop calculations.  The log-divergent parts of these coefficients, of our main interest, are given as
\begin{eqnarray}
a &=&  -3 \Delta  \nonumber \\
b &=& 3 \cdot \frac{m_{0u}^2+m_{0h}^2+m_{03}^2}{m_{0l}^2}\cdot \Delta
\nonumber \\
c &=& a,
\end{eqnarray}
where $ \Delta = i \int \frac{d^{4}k}{(2\pi)^4}\frac{1}{(k^2+M_{GUT}^2)^2}$ with a generic GUT mass scale $M_{GUT}$. $m_{0l}^2$, $m_{0u}^2$, $m_{0h}^2$ are SUSY breaking masses for right-handed charged slepton, right-handed up-type squarks, and 5-plet Higgs, respectively, and $m_{03}$ is SUSY breaking trilinear coupling of $\tilde{l}\tilde{t}h$. The third relation $c = a$ is the consequence of U(1)$_{em}$ symmetry, or Ward identity, as discussed in the introduction.

When a specific relation $a = b = c$, which is equivalent to
\begin{equation}
m_{0l}^2 + m_{0u}^2 + m_{0h}^2 + m_{03}^2 = 0,
\end{equation}
holds, the SUSY breaking term $(\delta_{ij} + b H_{ij}) m_{0l}^2 $ has the same structure as other terms with coefficients $a$ and $c$, and these terms (and the kinetic term for charged leptons) can be diagonalized and rescaled simultaneously by a suitable wave-function renormalization. Thus  all LFV effects go away in $L_{rel}$, and the log-divergence $\Delta$ does not remain. Let us note the photino vertex does not have any LFV either, as both lepton and slepton mass matrices are simultaneously diagonalized.  This disappearance of LFV can also be checked diagramatically. In fact,  if $a = b$ holds the sum of ``external-leg correction" diagrams, picking up the effect of either $a$ or  $b$ at the first order of $m_{0l}^2$, just disappears (see Fig.1 ).
%%%%%%%%%%%%%%%%%%%%%%%%%%%%%%%%%%%%%%%%%%%%%%%%%%%
\begin{center}
\begin{picture}(300,70)(0,0)
%%%
\DashLine(5,35)(32,35){2.5}
\DashLine(38,35)(62,35){2.5}
\DashLine(68,35)(95,35){2.5}
\DashLine(115,35)(142,35){2.5}
\DashLine(148,35)(195,35){2.5}
\Text(215,35)[l]{${\ldots \ldots }$}
%\DashLine(215,35)(275,35){2.5}
\Text(105,35)[]{${+}$}
\Text(205,35)[]{${+}$}
\Text(285,35)[]{${=}$}
\Text(300,35)[]{\Large ${0}$}
%\Photon(280,25)(305,25){1}{5}
%\Photon(280,23)(305,23){1}{5}
\CArc(35,35)(3,0,360)
\Line(33,37)(37,33)
\Line(33,33)(37,37)
\CArc(65,35)(3,0,360)
\Line(63,37)(67,33)
\Line(63,33)(67,37)
\CArc(145,35)(3,0,360)
\Line(143,37)(147,33)
\Line(143,33)(147,37)
\Photon(80,35)(80,5){3}{4}
\Photon(170,35)(170,5){3}{4}
\Text(15,40)[]{${\tilde{\mu}}$}
\Text(50,40)[]{${\tilde{e}}$}
\Text(75,40)[]{${\tilde{e}}$}
\Text(85,40)[]{${\tilde{e}}$}
\Text(125,40)[]{${\tilde{\mu}}$}
\Text(160,40)[]{${\tilde{e}}$}
\Text(185,40)[]{${\tilde{e}}$}
\Text(35,25)[]{${a p^2}$}
\Text(65,25)[]{${m^2_{0L}}$}
\Text(145,25)[]{${b m^2_{0L}}$}
%\Text(245,42)[]{(?)}
\Text(90,20)[]{${\gamma}$}
\Text(180,20)[]{${\gamma}$}
\Text(50,2)[]{\sl (a)}
\Text(155,2)[]{\sl (b)}
%\Text(245,2)[]{\sl (c)}
%%%

\end{picture} \\{{\sl Fig.1 }{\rm : Diagrams contributing to the sub-process 
${\tilde{\mu} \to e \tilde{\gamma}}$ of ${\mu \to e \gamma}$ at ${{\cal O} 
\left( M^2_{\rm SUSY} \right)}$. When $a = b$ holds the sum of these 
diagrams vanishes. }}
\end{center}
%%%%%%%%%%%%%%%%%%%%%%%%%%%%%%%%%%%%%%%%%%%%%%%%%%%%%%%%%%%%%%%%%%%%%%%%%%%  
Such cancellation mechanism does not seem to have been addressed  in previous literature. Since right-handed charged lepton and right-handed up-type quark super-multiplets both belong to 10-plet repr. of SU(5), if we  write $m_{0l}^2 = m_{0u}^2 = m_{10}^2$ and $m_{0h}^2 = m_{5}^2$, and neglect the trilinear coupling, the condition quoted above reduces to $10\cdot m_{10}^2 + 5\cdot m_{5}^2 = Str(M^2) = 0$, where ``super-trace"
 $Str$ is over a anomaly free set of matter multiplets. This might suggest some theory which might have been hidden behind and provides a scheme of natural lepton flavour conservation.

In SUGRA-inspired SUSY GUT, however, the above condition is clearly not met, and the logarithmically divergent LFV effect seems to remain. How should we interprete the remaining logarithmic divergence ? The LFV SUSY breaking  slepton mass-squared  is the coefficient of a gauge invariant $d = 2$ operator and should belong to $L_{rel}$, though we did not include it at the tree level. This is why the divergence appeared after the quantum correction. The divergence  should be removed by the introduction of counterterms and is subject to a renormalization condition. Thus we are forced to make a conclusion that unfortunately the logarithmic correction cannot be taken as a prediction of the theory, at least as far as we work in the framework of SUSY GUT, as a renormalizable theory (calculable predictions of a theory should come from the finite quantum corrections to irrelevant operators). Let us note that ignoring the LFV mass-squared  operators, $m_{LFV}^2 \ \tilde{e}_R^{\ast}\tilde{\mu}_R$, $ m_{LFV}^2  
(|\tilde{\mu}_R|^2 - |\tilde{e}_R|^2)$, etc.,  does not enhance any symmetry of the theory, since SUSY has already been explicitly broken by the flavour independent SUSY breaking masses and also the flavour symmetry has been broken by the large top Yukawa coupling.  In other words, there is no symmetry in the original lagrangian which guarantee the smallness of the lepton flavour violation. We thus might have to say that {\it ``the theory does not lead to natural lepton flavour conservation."}

One may wonder the situation discussed in Ref.\cite{Barbieri} that SUSY breaking slepton masses, whose boundary condition is set to be flavour independent at $\Lambda = M_{pl}$, get logarithmic LFV corrections by the renormalization group effect is similar to the well-known evolution of gauge couplings in ordinary SU(5) GUT, where three gauge couplings for SU(3), SU(2) and U(1), set to be all equal at $M_{GUT}$, deviate from each other in lower energies by renormalization group effect. Actually the situation is quite different in these two cases. In the case of the evolution of gauge couplings the universal coupling at $M_{GUT}$ is naturally guaranteed by the symmetry of the theory, i.e. by SU(5). Thus the splitting of gauge couplings comes not from the correction to the $L_{rel}$ but from the appearance of a new irrelevant operator with adjoint Higgs field included, in order to trace the spontaneous breaking of SU(5). Thus the splitting is genuine prediction of the theory, and is independent of the choice 
of the cutoff $\Lambda$. On the other hand, in the case under consideration the log-correction depends on the cutoff $\Lambda$ and further there will be no reason to expect that all SUSY breaking slepton masses evolve equally above $M_{pl}$ (, though we are not sure what the ``above $M_{pl}$" means), even if they are once unified at $M_{pl}$. We should note that flavour symmetry is hardly broken by large top Yukawa coupling, while SU(5) symmetry is softly broken by the VEV of the adjoint Higgs.

Finally we will ask  a question whether there is some chance to get the logarithmic non-decoupling  LFV effect as a natural prediction of some renormalizable theory. More specifically, let us think of a GUT model with ``gauge mediated SUSY breaking" as a typical example. At first glance the situation is very similar to that of the SUGRA theory with hidden sector. However, in the ``gauge-mediated" scenario SUSY breaking in the observable sector appears only at the quantum level, while in SUGRA the SUSY breaking is transmitted to the observable sector already at the tree level via non-renormalizable (super-)gravitational interaction. As the result, we will not get the non-decoupling effect in the renormalizable theory: we cannot expect to get a log-divergent quantum correction to the LFV slepton masses in this case. The reason is that the absence of the LFV slepton masses in the lagrangian enhances supersymmetry in the observable sector, in contrast to the case of SUGRA where the absence does not enhance SUSY 
as there are universal SUSY breaking masses already at the tree lagrangian.  Thus as long as the theory is renormalizable, there will not appear any log-divergent correction to the masses. The LFV slepton masses, therefore,  should be described by a higher dimensional ($d >4$)irrelevant operator whose coefficient is finite, and is expected to be suppressed at least by $\frac{M_{SUSY}^2} {M_{GUT}^2}$, with $M_{SUSY}$ being SUSY breaking mass scale in the observable sector. Hence the rate of resultant $\mu \rightarrow e\gamma$ is anticipated to be outrageously suppressed. It is interesting to note that the aforementioned condition in order to cancel the log-divergence is trivially satisfied in this type of theories, as SUSY breaking masses, $m_{0l}^2$ etc., are all absent at the tree level.

\noindent {\bf Acknowledgment}

  The author would like to thank  T. Inami  for very fruitful discussions.  Thanks are also due to N. Haba for  useful and informative arguments on the subject of gauge mediated SUSY breaking scenario. This work has been supported by Grant-in-Aid for Scientific Research (09640361) from the Ministry of Education, Science and Calture, Japan.

\end{document}